\documentclass[graybox]{svmult}

\usepackage{mathptmx}       % selects Times Roman as basic font
\usepackage{helvet}         % selects Helvetica as sans-serif font
\usepackage{courier}        % selects Courier as typewriter font
\usepackage{type1cm}        % activate if the above 3 fonts are
\usepackage{makeidx}         % allows index generation
\usepackage{graphicx}        % standard LaTeX graphics tool
\usepackage{multicol}        % used for the two-column index
\usepackage[bottom]{footmisc}% places footnotes at page bottom

\newcommand{\beq}{\begin{equation}}
\newcommand{\eeq}{\end{equation}}
\newcommand{\beqa}{\begin{eqnarray}}
\newcommand{\eeqa}{\end{eqnarray}}
\newcommand{\nn}{\nonumber \\}

\def \b {\beta}
\def \disk {\mathrm{disk}}
\def \e {\mathrm{e}}

\def \la {\langle}
\def \ra {\rangle}

\def \t {\tau}

\def \CFT {\mathrm{CFT}}

\def \H {{\mathcal H}}

\def \df {\stackrel{\mathrm{def}}{=} \ }
\def \el {\mathrm{el}}
\def \imb {\mathrm{imb}}

\def \z {\zeta}

\def \D {\Delta}
\def \I {{\mathbb I}}

\def \Im {\mathrm{Im} \, }

\def \H {{\mathcal H}}
\def \uu {{\widehat{u(1)}}}

\begin{document}

\title*{Thermopower in the Coulomb blockade regime for Laughlin quantum dots}
\author{Lachezar S. Georgiev}
\institute{Lachezar S. Georgiev \at Institute for Nuclear Research and Nuclear Energy, Bulgarian Academy of Sciences, 
72 Tsarigradsko Chaussee, 1784 Sofia, Bulgaria, \email{lgeorg@inrne.bas.bg}}
\maketitle

\abstract{
Using the conformal field theory partition function of a Coulomb-blockaded quantum dot, 
constructed by two quantum point contacts in a Laughlin quantum Hall bar, we derive the 
finite-temperature thermodynamic expression for the thermopower in the linear-response regime.
The low-temperature results for the thermopower are compared to those for the conductance 
and their capability to  reveal the structure of the single-electron spectrum in the quantum 
dot is analyzed.
}
%%%%%%%%%%%%%%%%%%%%%%%%%%%%%%%%%%%%%%%%%%%%%%%%%%%%%%%%
\section{What are Quantum Dots and why study them?}
\label{sec:QD}
%%%%%%%%%%%%%%%%%%%%%%%%%%%%%%%%%%%%%%%%%%%%%%%%%%%%%%%%
Quantum dots (QD) are mesoscopic conducting islands of two-dimensional (incompressible) electron gas constructed 
on the metal-oxide-semiconductor interface in a typical field-effect transistor \cite{kouwenhoven,matveev-LNP}. 
The semiconductor bar contains a small number of bulk charge carriers (electrons or holes) which are pushed out to 
an overlaying oxide insulator layer by means of 
electric field perpendicular to the interface surface, creating in this way a two-dimensional film of strongly 
correlated electrons with a finite geometry realized by a confining potential. Under appropriate conditions 
(low temperature, high perpendicular magnetic fields in a high-mobility semiconductor samples) the strongly 
correlated electron gas can be found to be in the  quantum Hall regime (integer or fractional) and for simplicity we 
will think of it as a two-dimensional droplet of quantum Hall liquid with disk shape whose dynamics is concentrated on the 
one-dimensional edge which is a circle. 

The QDs have a number of interesting properties and are essential part of the so called Single-electron transistors 
(SET) which explains why they have been the subject of intense research in recent years. 
Because of the small size of the QDs (typical circumference of several $\mu m$) and its isolation form the rest of  the 
system (only small tunneling is considered), QDs are almost closed quantum systems with a discrete energy spectrum at very 
low temperatures, which make them similar to large artificial atoms in which one can investigate both fundamental concepts of 
quantum theory and important application aspects of nanoelectronics as well as transcend the cutting-edge 
research-and-development perspectives for the implementation  of  quantum computers and quantum information processing.

The incompressible fractional quantum Hall liquids have been successfully described by two-dimensional 
rational  conformal field theories \cite{CFT-book} (CFT) governing  the dynamics of their edge excitations \cite{cz}. 
In this contribution we will show how one can use the CFT for QDs, 
realized inside of quantum Hall bar corresponding to the  $\nu_H=1/m$ Laughlin state,  to 
calculate observable thermodynamic characteristics of the QDs, such as the tunneling conductance and thermopower.
%%%%%%%%%%%%%%%%%%%%%%%%%%%%%%%%%%%%%%%%%%%%%%%%%%%%%%%%
\section{Quantum dots and Single-electron transistors}
\label{sec:SET}
\begin{figure}[htb]
\centering
\includegraphics[bb=0 300 600 530,clip,width=\textwidth]{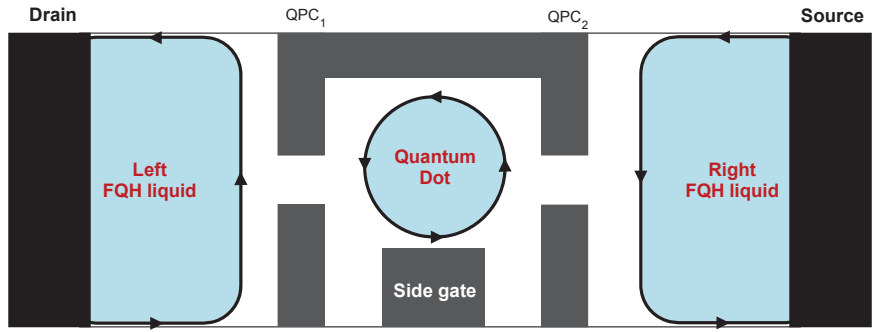}
\caption{Single-electron transistor realized by two quantum-point contacts (QPC$_1$ and QPC$_2$) inside of
 $\nu_H=1/m$ Laughlin FQH state. The arrows show the direction of the propagation of the edge modes. 
Only electrons can tunnel between the left and right FQH liquids and the QD under 
appropriate conditions. \label{fig:SET}}
\end{figure}

%%%%%%%%%%%%%%%%%%%%%%%%%%%%%%%%%%%%%%%%%%%%%%%%%%%%%%%%
When a QD is equipped with drain and source gates, as shown on Fig.~\ref{fig:SET},
%%%%%%%%%%%%%%%%%%%%%%%%%%%%%%%%%%%%%%%%%%%%%%%%%%%%%%%%
%%%%%%%%%%%%%%%%%%%%%%%%%%%%%%%%%%%%%%%%%%%%%%%%%%%%%
 by applying a drain-source voltage one could in 
principle transfer electrons from the left FQH liquid to the QD and then to the right FQH liquid. However, a 
tunneling electron from left to the QD must overcome the Coulomb charging energy  $e^2/2C$, associated 
with adding one extra electron to the QD, where $C$ is the total capacitance of the QD. When the QD is 
small so is $C$ 
and this  Coulomb charging energy could be large, so that at low temperature $k_BT\ll e^2/C$ and small bias
the electron transfer is blocked. This is called the Coulomb blockade
\cite{matveev-LNP,kouwenhoven,staring-CB}. Because we are interested in the small-bias regime, 
which can be treated by linear response, one way to lift the Coulomb blockade at small bias is 
to add a third electrode called the Side gate, see Fig.~\ref{fig:SET}. Then, by changing the gate voltage $V_g$ 
one can shift the discrete energy levels of the QD, still in the linear response regime, to align them with the 
Fermi levels of the left and right FQH liquids and when this happens one electron can tunnel from left to the 
right through the QD. Since the electrons tunnel one-by-one with the variation of $V_g$ this three-gate QD 
construction is called a Single-electron transistor, see Fig.~\ref{fig:SET} for its scheme.

The QD in the SET is an almost closed quantum system of size from $0.1$ $\mu m$ to $1$ $\mu m$
with discrete single-electron energy levels of typical spacing $\Delta \epsilon =\hbar 2\pi v_F/L$, where $v_F$ 
is the Fermi velocity of the edge mode and $L$ is the circumference of the edge circle.
Only small tunneling is allowed between the leads and the QD, i.e., the tunneling conductances for $QPC_1$ 
and $QPC_2$ are much smaller than the conductance quantum: $G_{L/R} \ll e^2/h$, which guarantees that 
the single-particle energy levels in the QD remain discrete. At low temperature the number of electrons on the 
QD is quantized to be integer and can be computed as a derivative of the thermodynamic density of states 
with respect to the  chemical potential - here we can use the RCFT partition function as a thermodynamical 
Grand potential. Thus the QDs are very similar to large artificial atoms - almost 1000 times bigger than the 
average atoms, they are highly tunable, yet still purely  quantum systems! 
For example, one magnetic flux quantum in an atom requires magnetic field of the order of $10^6$ T, while for
 QDs the corresponding field is of order of $1$ T \cite{kouwenhoven}.
This makes QDs very convenient for verification of fundamental concepts of quantum theory as well as for 
quantum computation and information processing.

For small QD and small bias the charging effects  leading to the Coulomb blockade become important at low T 
such that  $k_B T  \ll e^2/C$.  The variation of the side gate voltage $V_g$  induces external
electric charge on the QD and creates charge imbalance between the QD and the side gate  which
 changes continuously the single-particle energies of the QD lifting in this way the CB \cite{kouwenhoven,staring-CB}.

Changing adiabatically the side gate voltage $V_g$ at small-bias tunneling,
 between the left- and right- FQH liquids and the QD, results in a precise QD level spectroscopy 
which can be treated analytically in the linear response  regime under the following conditions:
\begin{itemize}
{\setlength\itemindent{25pt} \item
low temperature $k_B T \ll e^2/C$ }
{\setlength\itemindent{25pt} \item 
low bias $V \ll e/C$ }
{\setlength\itemindent{25pt} \item
low QPC conductances $G_{L,R} \ll e^2/h$ }
\end{itemize}
Under these conditions the sequential tunneling of electrons one-by-one is dominating the cotunneling, which is a 
higher-order process associated with almost simultaneous virtual tunneling of pairs of 
electrons \cite{matveev-LNP}, that will not be considered here.
%%%%%%%%%%%%%%%%%%%%%%%%%%%%%%%%%%%%%%%%%%%%%%%%%%%%%%%%
%\begin{figure}[htb]
%\centering
%\includegraphics[bb=0 260 596 596,clip,width=\textwidth]{energies-1-3.pdf}
%\end{figure}
%%%%%%%%%%%%%%%%%%%%%%%%%%%%%%%%%%%%%%%%%%%%%%%%%%%%%%%%
\section{QD conductance--CFT spectroscopy}
\label{sec:spectroscopy}
%%%%%%%%%%%%%%%%%%%%%%%%%%%%%%%%%%%%%%%%%%%%%%%%%%%%%%%%
The tunneling conductance of the QD in the linear response regime can be computed  at low temperature 
from the Grand canonical partition function \cite{thermal}
\beq\label{Z_disk}
Z_{\disk}(\t,\z) = \mathrm{tr}_{ \H_{\mathrm{edge}}} \ \e^{-\beta (H_\CFT-\mu N_\el)} 
= \mathrm{tr}_{ \H_{\mathrm{edge}}} \ \e^{2\pi i \t (L_0 -c/24)} e^{2\pi i \z J_0}, 
\eeq
which describes the dynamics of the edge in terms of CFT assuming that the bulk of the QD is inert. 
In Eq.~(\ref{Z_disk}) we have denoted by $H_\CFT=\hbar\frac{2\pi v_F}{L} \left(L_0-\frac{c}{24}\right)$  the 
edge states' Hamiltonian, by $N_\el=-\sqrt{\nu_H} J_0$ the electron number operator on the edge, $L_0$ 
is the zero mode of the Virasoro stress-tensor \cite{CFT-book}, $J_0$ is the normalized zero mode of the 
$\uu$ current algebra \cite{CFT-book,LT9} and  $\nu_H$ denotes the FQH filling factor.
The trace in Eq.~(\ref{Z_disk})  is taken over the edge-states' Hilbert space $\H_{\mathrm{edge}}$  
whose structure might depend on the presence of quasiparticles in the bulk of the QD \cite{LT9}.

The modular parameters \cite{CFT-book} of the rational CFT are related to the temperature $T$ and 
chemical potential $\mu$ of the QD
\beq\label{parameters}
\t=i\pi\frac{T_0}{T}, \quad 
T_0=\frac{\hbar v_F }{\pi k_B L}, \quad \z= i\frac{1 }{2\pi  k_B T} \mu .
\eeq
The disk CFT partition function for the Grand canonical ensemble  in presence of AB flux $\phi$ can be expressed
in a compact way by shifting the chemical potential \cite{NPB-PF_k}
\beq \label{Z_AB}
\z \to \z +\phi \t, \quad Z_{\disk}^{\phi}(\t,\z ) \df
\mathrm{tr}_{ \H_{\mathrm{edge}}} \ \e^{-\beta (H_\CFT(\phi)-\mu N_\imb(\phi))} \equiv
Z_{\disk}(\t,\z +\phi\t) , 
\eeq
where $N_\imb(\phi) =N_\el - \nu_H \phi$ is the \textit{particle imbalance due to the gate voltage}, see the explanations after 
Eq.~(\ref{J_0s}) below; what we will need here is the last expression in Eq.~(\ref{Z_AB}).
The thermodynamic Grand potential on the edge is expressed in terms of the partition function as usual 
 \beq\label{Omega_phi}
\Omega_\phi(T,\mu)=-k_B T \ln Z_{\disk}^{\phi}(\t,\z).
\eeq
The edge conductance has been shown  to be proportional 
to the  derivative of the thermodynamic density of states with respect to the chemical potential  \cite{thermal}, i.e.
\beq \label{G_is}
G_{\mathrm{is}} (\phi)=\frac{e^2}{h}
\left( \nu_H +\frac{1}{2\pi^2} \left(\frac{T}{T_0} \right)\frac{\partial^2 }{\partial \phi^2}  \ln Z_{\phi}(T,0)\right) .
\eeq
The conductance for the $\nu=1/3$ Laughlin QD, computed by Eq.~ (\ref{G_is}) from the partition function (\ref{K}) 
 given in the next section with $l=0$ at temperature $T=T_0$,
shows vast regions in which it is zero (CB valleys) and sharp peaks at values $\phi_i=3/2 + 3i$, $i=0,\pm 1, \pm 2, \ldots$ 
as shown in  Fig.~\ref{fig:N-average}.
%%%%%%%%%%%%%%%%%%%%%%%%%%%%%%%%%%%%%%%%%%%%%%%%%%%%%%%%
\section{The Laughlin QD partition function}
\label{sec:PF}
%%%%%%%%%%%%%%%%%%%%%%%%%%%%%%%%%%%%%%%%%%%%%%%%%%%%%%%%
The grand partition function for the edge of a QD in the $\nu_H=1/m$ Laughlin FQH state can be written as  
\beq\label{K}
 K_{l}(\t,\z; m) = \frac{\mathrm{CZ}}{\eta(\t)} \sum_{n=-\infty}^{\infty} q^{\frac{m}{2}\left(n+\frac{l}{m}\right)^2} 
\e^{2\pi i \z \left(n+\frac{l}{m}\right)} ,
\eeq
where $q=\e^{-\beta\D\varepsilon}=\e^{2\pi i \t}$ with $\b=(k_B T)^{-1}$ and 
$\D \varepsilon= \hbar\frac{2\pi v_F}{L}$. The index of the $K$-function 
$l =-(m-1)/2, \ldots, (m-1)/2$ ($m$ must be an odd integer) 
corresponds to a Hilbert space $\H_l$  with quasiparticles in the bulk \cite{LT9} with electric charge $l/m$.
The Dedekind function $\eta$  and the Cappelli--Zemba factor \cite{cz}
are
\[
 \eta(\t)=q^{1/24}\prod_{n=1}^\infty (1-q^n), \quad \mathrm{CZ}=\e^{-\pi\nu_H\frac{(\Im \z)^2}{\Im\t}},
\]
however, for our purposes they would be unimportant since we would set $\z=0$ at the end \cite{NPB-PF_k,thermal}.
%%%%%%%%%%%%%%%%%%%%%%%%%%%%%%%%%%%%%%%%%%%%%%%%%%%%%%%%
\section{Thermopower: a finer spectroscopic tool}
\label{sec:TP}
\begin{figure}[htb]
\centering
\includegraphics[bb=40 10 570 390,clip,width=12cm]{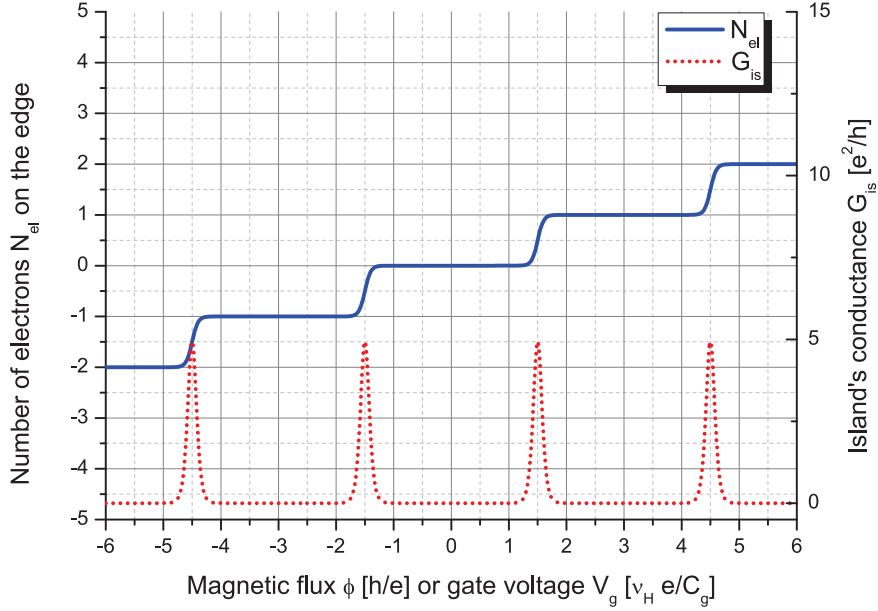}
\caption{Electron number average $N_\el$ on the edge and Coulomb blockade conductance $G_\mathrm{is}$ 
for the $\nu_H=1/3$ Laughlin island without bulk quasiparticles 
as a function of the gate voltage at temperature $T=T_0$.\label{fig:N-average}}
\end{figure}

%%%%%%%%%%%%%%%%%%%%%%%%%%%%%%%%%%%%%%%%%%%%%%%%%%%%%%%%
The thermopower $S$, known also as the Seebeck coefficient, is the potential difference $V$ between the leads of 
the SET when the two leads are at different  temperature  $T_R$ and $T_L$, assuming that the difference is small 
$\Delta T=T_R -T_L\ll T_L$, under the condition that the current $I$ between the leads is zero \cite{matveev-LNP} .
Usually thermopower is expressed as the ratio of the thermal conductance $G_T$ and electric conductance $G$, i.e.,  
$S=G_T/G$, however, this expression is not appropriate for SETs because $G=0=G_T$, 
while their ratio is finite, in vast
intervals of flux (in the CB valleys), see Fig.~\ref{fig:N-average}. 
Fortunately, there is an alternative expression in terms of the average  energy
 $\la \varepsilon \ra$ of the electrons tunneling through the QD \cite{matveev-LNP}
\[
S \equiv  \left. -\lim_{\Delta T \to 0} \frac{V}{\Delta T}\right|_{I=0}=-\frac{\la \varepsilon \ra}{eT}.
\]
where $T=T_L+\Delta T/2$ is the temperature of the QD.

The average tunneling energy could be computed thermodynamically using as thermodynamical potential 
the rational CFT partition function for the FQH edge of the QD. To this end we notice that due to energy conservation
in single-electron tunneling the average tunneling energy is simply the difference between the  total thermodynamic 
average energy of the QD
with $N+1$ and $N$ electrons at the same temperature $T$ and AB flux $\phi$ (respectively, gate voltage $V_g$)
divided by the difference in the electron numbers of the QD as a function of $\phi$
\beq \label{average}
\la \varepsilon \ra^{\phi}_{\beta,\mu_N} = 
\frac{E^{\beta,\mu_{N+1}}_{\mathrm{QD}}(\phi)-  E^{\beta,\mu_{N}}_{\mathrm{QD}}(\phi) }
{N_{\mathrm{QD}}^{\beta,\mu_{N+1}}(\phi) -  N_{\mathrm{QD}}^{\beta,\mu_N}(\phi)}.
\eeq
Because we are working within the Grand canonical ensemble, the total energy of the QD with $N$ electrons 
requires the chemical potential $\mu_N$ to be determined. 
It is defined as the chemical potential for which the average of the particle 
number operator is equal to the number  $N$ at zero gate voltage (AB flux)
\beq \label{mu_N}
\nu_H\left(\frac{\mu_N}{\Delta \epsilon} +\phi\right) -\frac{\partial \Omega_\phi(\beta,\mu_N)}{\partial \phi}= N  .
\eeq
The total energy of an $N$-electron QD within the \textit{Constant Interaction model} \cite{kouwenhoven} is
\beq \label{E_N}
E^{\beta,\mu_{N}}_{\mathrm{QD}}(\phi) =\sum_{i=1}^{N_0} E_i (B)+ \la H_{\mathrm{CFT}}(\phi)\ra_{\beta,\mu_{N}} 
 +U(N),
\eeq
where $N_0$ is the number of electrons in the bulk of the QD and $N-N_0=N_\el$ is the number of electrons on the edge,
 $E_i(B)$, $i=1, \ldots , N_0$, are the energies of the occupied single-electron states in the bulk of the QD,
the expectation value $\la \cdots \ra_{\beta,\mu}$ is the Grand canonical average of the Hamiltonian 
$H_{\mathrm{CFT}}$ on the edge,
and $U(N)$  is the ($B$-independent) electrostatic energy of the QD, including the contribution due to the 
gate voltage $V_g$ is (see Eq.~(1) in \cite{kouwenhoven})
\beq \label{U_N}
U(N)= \frac{\left[e(N-N_0)-C_gV_g\right]^2}{2C}  ,
\eeq
where $N=N_0$ for $V_g=0$.
The total capacitance $C=C_g+C_1+C_2$, where $C_g$ is the capacitance of the side gate, 
$C_1$ and $C_2$ are the capacitances of the two QPCs, 
is assumed independent of $N$ and this assumption a characteristic for the 
Constant Interaction model \cite{kouwenhoven}. 
Within this model the energies $E_i$ depend on the 
magnetic field $B$ and on the gate voltage $V_g$, but not on $N$ \cite{staring-CB}. In the case of
a FQH island we know that the variation of $V_g$ modifies also the single-electron energies on the edge
\cite{Stern-CB-RR,CB,stern-CB-RR-PRB,thermal}
due to a variation of the CB island's area $A$, producing a variation of the AB flux $\phi$. 
Because the variation of the  gate voltage $V_g$ induces (continuously varying) ``external charge" 
$e N_g=C_g V_g$ on the edge,  it is equivalent to the AB flux-induced variation of the particle number 
$N_\phi=\nu_H \phi$, so that 
we can take into account the subtler effects of the gate voltage on the edge energies 
$\la H_{\mathrm{CFT}}(\phi)\ra_{\beta,\mu_{N}}$ by introducing AB flux $\phi$ determined 
from~\footnote{for a one-dimensional circular edge all thermodynamic quantities 
depend on the magnetic flux not on the magnetic filed itself. Thus, the flux of the constant $B$ has the same effect
on the partition function as the singular AB flux, which is however, easier to take into account analytically \cite{NPB-PF_k}.}
 \beq \label{flux-voltage}
\frac{C_gV_g}{e}\equiv \nu_H\phi, \quad \phi=\frac{e}{h}\left( A-A_0\right)B,
\eeq
where $A_0$ is the area of the CB island at $V_g=0$.
Therefore, when we speak about Coulomb blockade caused by a variation of the AB flux $\phi$ we actually mean a 
variation of the gate voltage $V_g$ determined from (\ref{flux-voltage}).
It is worth stressing that the electron number $N_\el$ on the QD is quantized to be integer, while ``particle number imbalance" 
$N_\imb=(N-N_0)-C_gV_g/e$, between the QD and the side gate, changes continuously when the gate voltage $V_g$ is varied 
\cite{kouwenhoven,staring-CB}. It is also interesting to mention that according to (\ref{flux-voltage}) the AB flux distance 
between two neighboring CB peaks is $\Delta\phi=\nu_H^{-1}$ because then $\Delta N_\phi=1$ so that an entire additional 
electron can be transferred through the QD. It corresponds to gate voltage periodicity between CB 
peaks equal to $e\Delta V_g = (1/\alpha_g) (e^2/C)$, where $\alpha_g=C_g/C$ is called the gate's lever arm \cite{kouwenhoven}.

Using the AB flux instead of the gate voltage like in Eq.~(\ref{flux-voltage}) is convenient because the flux 
can be interpreted mathematically as a continuous twisting of the $\widehat{u(1)}$ charge of the underlying chiral 
algebra  \cite{NPB-PF_k,CFT-book}, which is technically similar to the rational (orbifold) twisting of $\widehat{u(1)}$ 
current  \cite{kt}, i.e., its zero mode is modified by
\beq \label{J_0s}
J_0 \to \pi_\beta(J_0) =J_0-\beta \quad \mathrm{with} \quad \beta=-\sqrt{\nu_H}\phi.
\eeq
Then the average of the twisted  electric $\widehat{u(1)}$  current $\pi_\beta(J_0^\el)\equiv \sqrt{\nu_H}\pi_\beta(J_0)$ is 
proportional to the thermodynamic derivative of the Grand potential
$\partial \Omega_\phi/\partial \phi = \la \pi_\phi(J^{\el}_0)\ra$ whose physical meaning is the electrostatic 
charge imbalance 
between the CB island and the gate arising due to the gate voltage. The untwisted  $\widehat{u(1)}$ charge, which 
is proportional to the electron number on the edge 
$J^{\el}_0=\sqrt{\nu_H} J_0=-N_\el$, is according to (\ref{J_0s})
$\la J^{\el}_0\ra=\la \pi_\phi(J^{\el}_0)\ra -\nu_H\phi $ and this is equivalent to the following Grand canonical 
thermal average of the electron particle number on the edge, which is illustrated in Fig.~\ref{fig:N-average}
for the $\nu_H=1/3$ Laughlin state without quasiparticles in the bulk
\beqa \label{Nel-average}
\la N_\el (\phi)\ra_{\beta,\mu_N} &=& {-\frac{\partial \Omega_\phi(\beta,\mu_N)}{\partial \phi} } + {\nu_H\phi}
 +\nu_H\left(\frac{\mu_N}{\Delta \epsilon} \right) \nn
 &=&\nu_H\left(\phi+ \frac{\mu_N}{\Delta \epsilon} \right) +\frac{1}{2\pi^2} \left(\frac{T}{T_0}\right) \frac{\partial }{\partial \phi}
 \ln Z_\phi(T,\mu_N)
\eeqa
%%%%%%%%%%%%%%%%%%%%%%%%%%%%%%%%%%%%%%%%%%%%%%%%%%%%%%%%
%%%%%%%%%%%%%%%%%%%%%%%%%%%%%%%%%%%%%%%%%%%%%%%%%%%%%%%%
%%%%%%%%%%%%%%%%%%%%%%%%%%%%%%%%%%%%%%%%%%%%%%%%%%%%%%%%
\section{Average tunneling energy}
\label{sec:average}
%%%%%%%%%%%%%%%%%%%%%%%%%%%%%%%%%%%%%%%%%%%%%%%%%%%%%%%%
Taking into account Eqs.~(\ref{average}) and (\ref{E_N}), and neglecting the electrostatic energy $U(N)$ for large CB islands
as in Ref.~\cite{viola-stern},  we can compute the thermodynamic average 
energy of a single electron tunneling to the QD with $N$ electrons by
\beq\label{average2}
\la \varepsilon \ra^{\phi}_{\beta,\mu_N} =
\frac{\la H_{\mathrm{CFT}}(\phi)\ra_{\beta,\mu_{N+1}} - \la H_{\mathrm{CFT}}(\phi)\ra_{\beta,\mu_N}}
{\la N_\el(\phi)\ra_{\beta,\mu_{N+1}} - \la N_\el(\phi)\ra_{\beta,\mu_N}}.
\eeq
Notice that  the first term in the r.h.s of Eq.~(\ref{E_N}) cancels, while the electrostatic energy $U(N)$
is subleading for large CB islands, which are of experimental  interest \cite{gurman-2-3,viola-stern}, and is omitted.

\begin{figure}[htb]
\centering
\includegraphics[bb=40 0 560 380,clip,width=11cm]{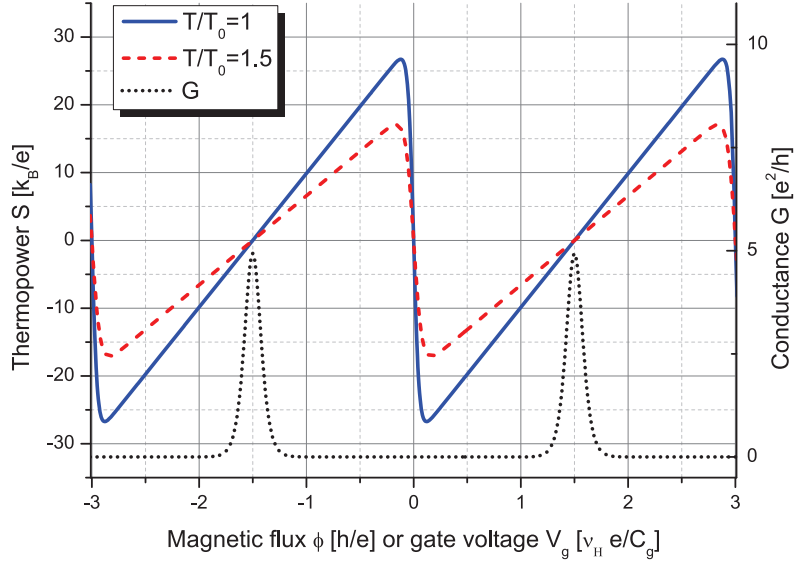}
\caption{Thermopower of the $\nu_H=1/m$ Laughlin  state with $m=3$ at temperatures $T=T_0$ and $T=1.5 T_0$. 
The conductance at $T=T_0$ is also shown on the right vertical scale. \label{fig:TP-G-L3-T10-15}}
\end{figure}

The average of the edge Hamiltonian is computed according to the standard formula for the Grand canonical ensemble \cite{kubo}
\beq \label{edge-average}
\la H_{\mathrm{CFT}}(\phi)\ra_{\beta,\mu_N} =\Omega_{\phi}(T,\mu_N)- T \frac{\partial \Omega_{\phi}(T,\mu_N)}{\partial T}  
 - \mu_N\frac{\partial  \Omega_{\phi}(T,\mu_N)}{\partial \mu}
\eeq
where $\Omega_{\phi}(T,\mu_N)$ is the Grand potential in presence of AB flux $\phi$ defined in (\ref{Omega_phi}).
Introducing the AB flux $\phi$ and chemical potential $\mu$ into the partition function (\ref{K}) according to (\ref{Z_AB}) 
and moving the $\phi$ and $\mu$ dependence into the index $l$ of (\ref{K}), see \cite{NPB-PF_k,thermal}, we obtain
(a factor independent of $\mu$ and $\phi$ is omitted)
\beq \label{K_phi}
 Z_{\phi}(T,\mu) =K_{\frac{\mu}{\Delta\epsilon}+\phi} (\tau,0;m)\propto
  \sum_{n=-\infty}^{\infty} 
q^{\frac{m}{2}\left(n+\frac{\mu/\Delta\epsilon+\phi}{m}\right)^2} .
 \eeq
The partition function (\ref{K_phi}) has a remarkable symmetry --  adding one electron to the ground state, 
which is equivalent to increasing the flux by $m$, does not change it, i.e., 
$Z_\phi(T,\mu^{\mathrm{GS}}_{N+1})=Z_{\phi+m}(T,\mu^{\mathrm{GS}}_{N})=Z_\phi(T,\mu^{\mathrm{GS}}_{N})$, implying
$\Omega_{\phi}(T,\mu^{\mathrm{GS}}_{N+1}) = \Omega_{\phi}(T,\mu^{\mathrm{GS}}_{N})$ and 
\beq\label{identity}
\frac{\partial  \Omega_{\phi}(T,\mu^{\mathrm{GS}}_{N+1})}{\partial T} = \frac{\partial  \Omega_{\phi}(T,\mu^{\mathrm{GS}}_{N})}{\partial T} ,\quad
\frac{\partial  \Omega_{\phi}(T,\mu^{\mathrm{GS}}_{N+1})}{\partial \mu} = \frac{\partial  \Omega_{\phi}(T,\mu^{\mathrm{GS}}_{N})}{\partial \mu} .
\eeq
Using the symmetry (\ref{identity}) we can find the difference between the ground-states 
chemical potentials of the QD with $N$ and $N+1$ 
electrons. Indeed, writing Eq.~(\ref{mu_N}) for $N$ and $N+1$ electrons 
\beqa
\nu_H\left(\frac{\mu^{\mathrm{GS}}_N}{\Delta \epsilon} +\phi\right) -
\frac{\partial \Omega_\phi(\beta,\mu^{\mathrm{GS}}_N)}{\partial \phi}&=& N \nn
\nu_H\left(\frac{\mu^{\mathrm{GS}}_{N+1}}{\Delta \epsilon} +\phi\right) -
\frac{\partial \Omega_\phi(\beta,\mu^{\mathrm{GS}}_{N+1})}{\partial \phi}&=& N+1  \nonumber
\eeqa
and subtracting them we obtain $\mu^{\mathrm{GS}}_{N+1}-\mu^{\mathrm{GS}}_N=m\Delta \epsilon$. 
This means that the chemical potentials $\mu^{\mathrm{GS}}_N$ and $\mu^{\mathrm{GS}}_{N+1}$ cannot be both set to $0$. 
Adjusting the chemical potential for $\phi=0$ to be in the middle between $\mu^{\mathrm{GS}}_N$ 
and $\mu^{\mathrm{GS}}_{N+1}$ (center of the CB valley), i.e., assuming  we obtain  
\[
\mu^{\mathrm{GS}}_N=-\frac{m}{2}\Delta\epsilon, \quad \mu^{\mathrm{GS}}_{N+1}=\frac{m}{2}\Delta\epsilon.
\]
These values of the chemical potentials determine the ground-state energies of the CB island with $N$ and $N+1$
electrons and their difference gives the addition energy characterizing the energy spacing of the CB conductance peaks.
However, for the calculation of the average tunneling energy (\ref{average2}) we need to find the difference between
 the energies of the $N$-th occupied single-particle state in the QD and the next available one, which is not the ground state
 with $N+1$ electrons. Instead, the next available single-particle state can be obtained from the last occupied state 
by increasing adiabatically the AB flux threading the edge by exactly one flux quantum. This is equivalent to
 increasing $\mu/\Delta\varepsilon$ by 1 so that the difference between the two chemical potentials is 
$\mu_{N+1}-\mu_N = \Delta\varepsilon$. Therefore, choosing again a symmetric setup so that $\mu_N + \mu_{N+1}=0$, 
we obtain
\beq \label{mu_N2}
\mu_N=-\frac{\Delta\epsilon}{2}, \quad \mu_{N+1}=\frac{\Delta\epsilon}{2}.
\eeq
Next, we can compute numerically the two edge energy averages (\ref{edge-average}) for a $\nu_H=1/3$ QD with $N$ and 
$N+1$ electrons with chemical potentials  (\ref{mu_N2}).   The plot of the thermopower for $T/T_0=1$ and $T/T_0=1.5$ 
and the conductance at $T/T_0=1$ are given in Fig.~\ref{fig:TP-G-L3-T10-15}.
%%%%%%%%%%%%%%%%%%%%%%%%%%%%%%%%%%%%%%%%%%%%%%%%%%%%%%%%
%%%%%%%%%%%%%%%%%%%%%%%%%%%%%%%%%%%%%%%%%%%%%%%%%%%%%%%%
The  plot of the thermopower has a sawtooth shape like that in metallic CB islands \cite{matveev-LNP}.
Also it is interesting to note that thermopower vanishes at the conductance peaks position in the same way as it 
does for metallic islands, expressing the fact that the energy difference between the QD with $N$ and $N+1$ electrons 
is zero at the maximum of the conductance peak. In the middle of the CB valleys the thermopower has sharp 
 jumps (discontinuous at $T=0$), expressing the particle-hole symmetry in the centers of the valleys \cite{matveev-LNP}.

%%%%%%%%%%%%%%%%%%%%%%%%%%%%%%%%%%%%%%%%%%%%%%%%%%%%%%%%
\section{Conclusion and perspectives}
\label{sec:concl}
%%%%%%%%%%%%%%%%%%%%%%%%%%%%%%%%%%%%%%%%%%%%%%%%%%%%%%%%
We have shown that the Constant Interaction model works fine for the Laughlin CB islands. 
Thermopower is non-zero in the CB valleys while the electric and thermal conductances are both zero.
The period of the thermopower is $\Delta\phi = m$
and its zeros correspond to the conductance peaks. 
Thermopower appears to be more sensitive to the neutral modes in the FQH liquid than the tunneling conductance 
 which explains why it is considered a finer spectroscopic tool. This could make thermopower
 an appropriate observable, which could distinguish between different FQH states with similar CB conductance patterns
 \cite{nayak-doppel-CB}, 
and therefore it would be interesting to apply this approach to FQH QDs with filling factors 
 $\nu_H=n_H/d_H$ for $n_H\ge 2$, especially  for non-Abelian FQH states. 
The sensitivity of the thermopower depends, however, on the relative sizes of 
the Coulomb charging energy and single-particle energies of the QD, which depend on the size and quality of the 
CB island. The experimental realization of CB islands in the fractional quantum Hall regime is challenging, however 
efforts have been made to measure the thermoelectric properties of such systems \cite{viola-stern}. 
For example, in a recent experiment these properties 
have been investigated for the $\nu_H=2/3$ FQH state \cite{gurman-2-3,viola-stern} which is similar to the $\nu_H=1/3$ 
Laughlin state but is expected to have a more complicated structure related to neutral modes.
%%%%%%%%%%%%%%%%%%%%%%%%%%%%%%%%%%%%%%%%%%%%%%%%%%%%%%%%%
\begin{acknowledgement}
I thank Andrea Cappelli, Guillermo Zemba and 
Bojko Bakalov for many helpful discussions.
This work has been partially supported by the Alexander von Humboldt Foundation under the Return Fellowship and 
Equipment Subsidies Programs and by the Bulgarian Science Fund under Contract No. DFNI-E 01/2.
\end{acknowledgement}
%
%\bibliography{my,TQC,FQHE,Z_k,CB}
\providecommand{\href}[2]{#2}\begingroup\raggedright\endgroup

\end{document}